\documentclass[twocolumn,prl,showpacs,preprintnumbers,amsmath,amssymb]{revtex4}

\usepackage{graphicx}
\usepackage{bm}

\begin{document}

\title{
Chiral skyrmions in thin magnetic films: 
new objects for magnetic storage technologies?
}
\author{N. S. Kiselev}
\author{ A. N.\ Bogdanov}
\author{R. Sch\"afer}
\author{U. K. R\"o\ss ler}
\email{u.roessler@ifw-dresden.de}

\affiliation{IFW Dresden, Postfach 270116, D-01171 Dresden, Germany}%

\begin{abstract}

{Axisymmetric magnetic lines of nanometer sizes (chiral \textit{vortices}
or \textit{skyrmions}) have been predicted to exist in a large group of
noncentrosymmetric crystals more than two decades ago.
Recently these magnetic textures have been directly 
observed in nanolayers of cubic helimagnets 
and monolayers of magnetic metals.
We develop a micromagnetic theory of chiral skyrmions in thin magnetic
layers for magnetic materials with intrinsic and induced chirality.
Such particle-like and stable micromagnetic objects 
can exist in broad ranges of applied magnetic fields including 
zero field.
Chiral skyrmions can be used as a new type of highly mobile nanoscale data carriers.
}
\end{abstract}

%

         
\maketitle

In magnetic materials with broken chiral symmetry
the structural handedness induces chiral
Dzyaloshinskii-Moriya (DM) couplings \cite{Dz64}
which stabilize two- and three-dimensional
localized structures with a fixed rotation
sense and nanometer sizes \cite{JETP89,Nature06}.
Originally, they have been described as chiral 
vortex-like configurations, but they 
are smooth and stable, topologically non-trivial 
magnetization configurations and, therefore, 
can be identified as \textit{skyrmions}
in the micromagnetic limit with constant 
magnetization modulus, $|\mathbf{M}|=\textrm{const}$ .
These skyrmions differ from 
other axisymmetric patterns
induced by external dipole-dipole forces
(bubble domains in nanolayers \cite{Kiselev10}
and magnetic vortices in magnetic nanodots \cite{Schneider00}).
Importantly, chiral skyrmions can
also arise in nanolayers of magnetic 
metals where they are stabilized by
surface/interface induced DM interactions \cite{PRL01}.
In common (centrosymmetric) magnetic
crystals such solitonic states are 
radially unstable and collapse spontaneously
under the influence of the applied field 
or anisotropy \cite{JETP89}.
This singles out magnets 
with intrinsic and induced DM interactions
into a particular class of  magnetic materials 
where nanoscale magnetic solitons exist \cite{Nature06}.

Recently, observations of
such chiral skyrmions have been reported
in nanolayers of noncentrosymmetric 
cubic ferromagnets (Fe,Co)Si and FeGe \cite{Yu10} 
and in monolayers of Fe with 
a strong surface induced DM coupling \cite{Heinze11}. 
This experimental break-through 
is not only an impressive demonstration of 
a unique phenomenon: static solitons
and formation of solitonic mesophases 
in a chiral condensed matter system \cite{Nature06}. 
These experiments also constitute new avenue for 
magnetic data storage and spintronics technologies.
Chiral skyrmions, as magnetic inhomogeneities 
localized into spots of a few nanometer,
can be freely created and manipulated, e.g., 
in extended layers of {\em magnetically soft materials}. 
These countable objects allow to fulfill a key
task in creating versatile magnetic patterns 
at the nanoscale.
\begin{figure}[!t]
\includegraphics[width=8.5cm]{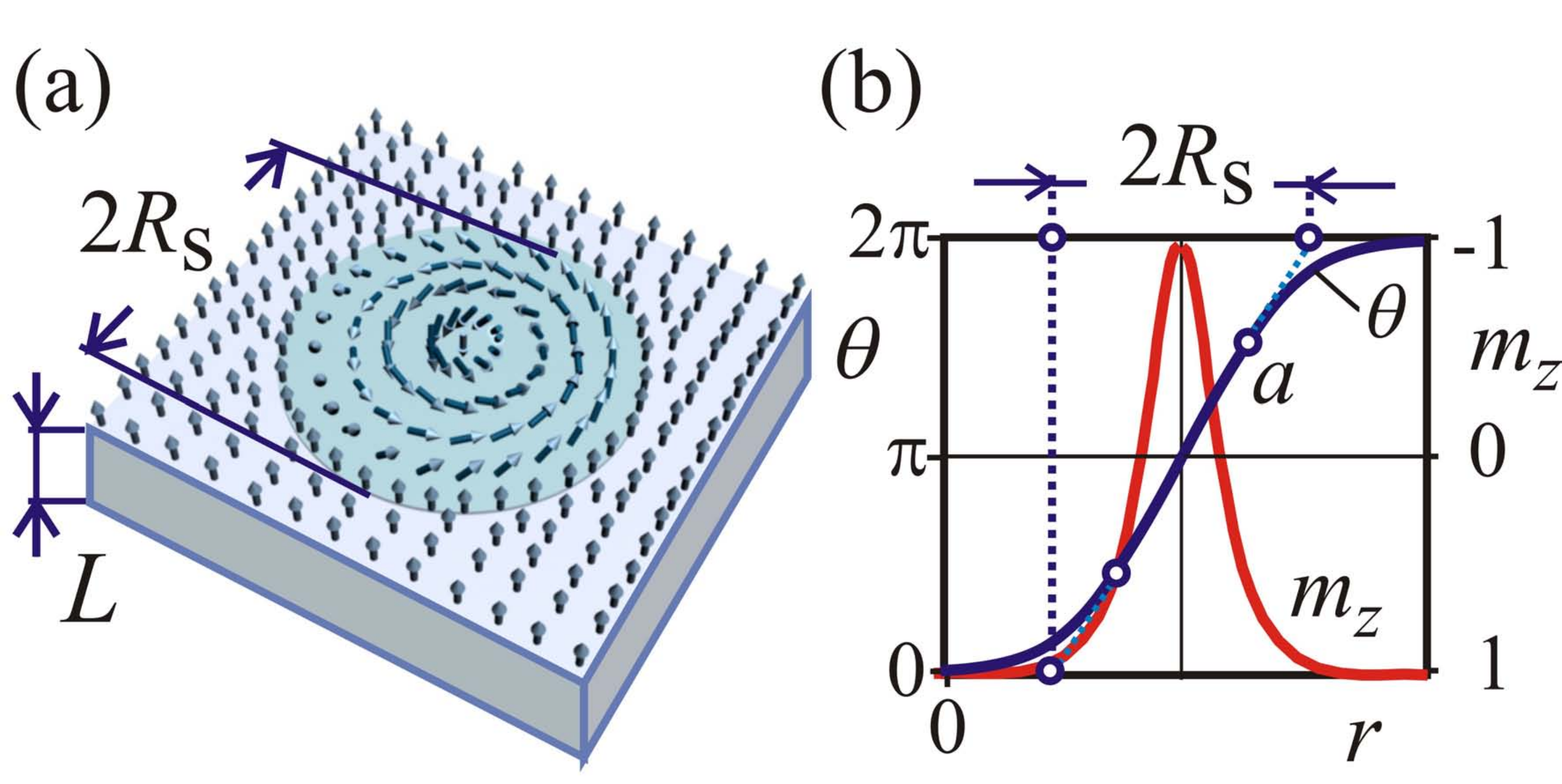}
\caption{ (a) Axisymmetric isolated skyrmion with the core diameter $2 R_s$
in a thin magnetic layer of thickness $L$. 
(b) The magnetization profile along the diameter cross-section
$\theta (r)$ and the perpendicular magnetization $m_z(r)$
indicate a strong localization of the skyrmion core;
$a$ ($r_0, \theta_0$) is the inflection point
and the core radius $R_s$ is derived from Eq.  (\ref{size})). 
\label{f1}
}
\end{figure} 

This letter formulates the basic micromagnetic 
theory for thin ferromagnetic layers with 
chiral DM couplings and presents equilibrium  
solutions  of isolated skyrmions applicable to a
broad range of material parameters.
This is the first step towards a detailed 
calculation of the properties and behavior 
of chiral isolated skyrmions in layer systems.

As a model we consider  a thin layer of
a uniaxial ferromagnet with intrinsic or induced 
Dzyaloshinskii-Moriya couplings.
The micromagnetic energy density of the layer
\cite{Dz64,Nature06}
\begin{eqnarray}
w = A (\mathbf{grad} \mathbf{M})^2 - \mathbf{M} \cdot \mathbf{H}
-K (\mathbf{M}\cdot \mathbf{n})^2 +w_d+ w_D
\label{density0}
\end{eqnarray}
includes exchange energy with stiffness $A$, 
uniaxial anisotropy with constant $K$ 
($\mathbf{n}$ is a unity vector perpendicular
to the layer surface), Zeeman, stray-field
($w_d$) and DM ($w_D$) energies \cite{Nature06,Hubert98}. 
The chiral DM couplings are written as 
antisymmetric differential forms
\begin{eqnarray}
 \Lambda_{ij}^{(k)}=M_i \frac{\partial M_j}{\partial x_k}
-M_j\frac{\partial M_i}{\partial x_k},
\label{Lifshitz}
\end{eqnarray}
that are known as  Lifshitz invariants \cite{Dz64}.
\begin{figure}[!t]
\includegraphics[width=8cm]{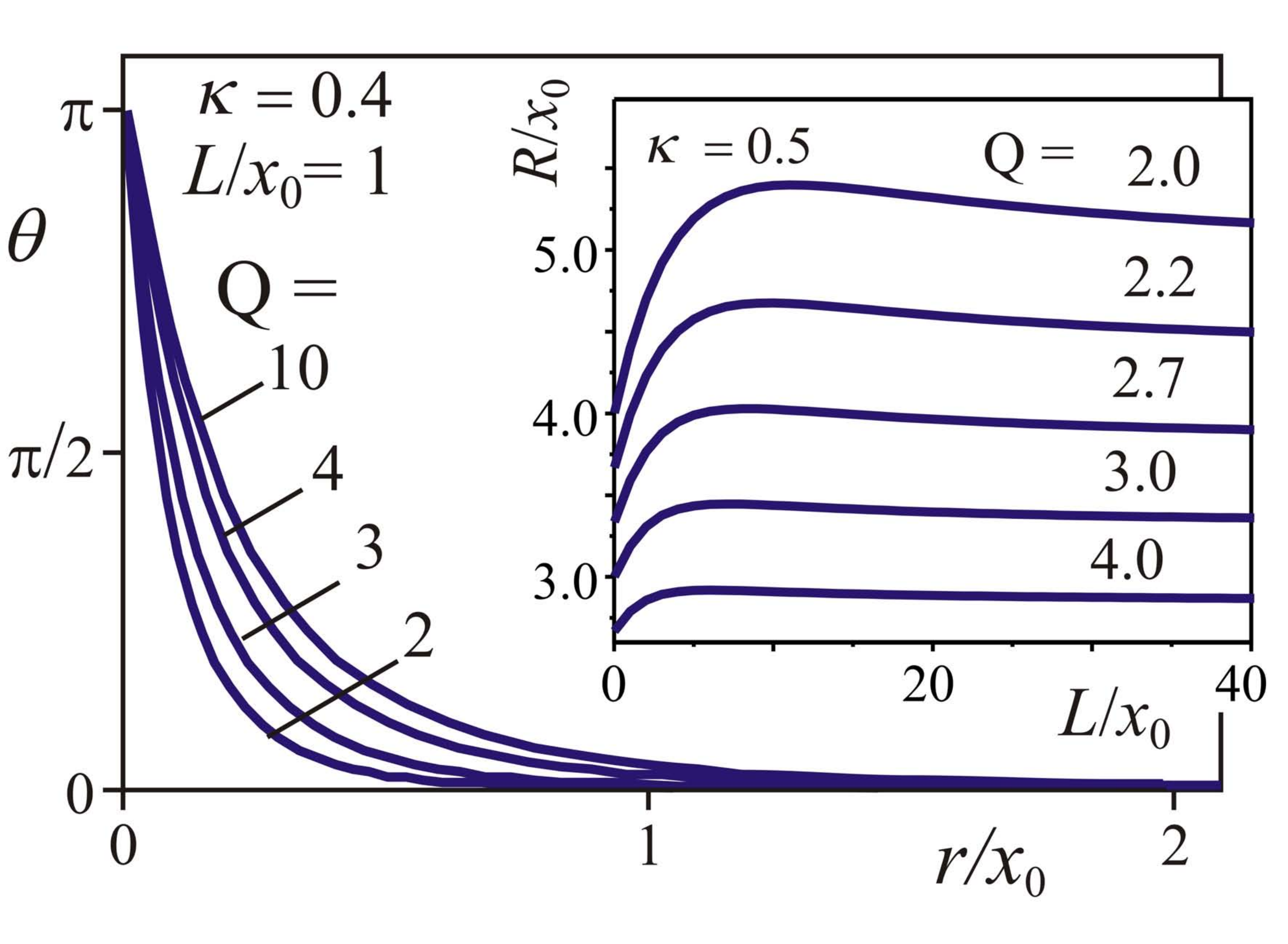}
\caption{ Equilibrium magnetization
profiles  $\theta (r)$ for model (\ref{energy})
with $H =0$, $\kappa = 0.4$, $L/x_0 =1$
and different values of $Q$ demonstrate
a strong localization of the skyrmion core.
Inset  shows  the reduced
skyrmion core sizes $R$ as a function
of the reduced layer thickness $L/x_0$
derived by minimization of $\Phi (R)$
(Eq. (\ref{phi})).
}
\label{profiles}
\end{figure}

Introducing spherical coordinates for the magnetization
$ \mathbf{M} = M (\sin \theta \cos \psi, \sin \theta \sin \psi,
\cos \theta$ ) and cylindrical coordinates  for the spatial
variable $\mathbf{r} = (r \cos \varphi, r \sin \varphi, z)$
one can show that the equations minimizing  functional (2)
include axisymmetric localized solutions of the
type $\theta (\rho)$, $\psi (\varphi)$ (Fig. \ref{f1}).
The solutions $\psi (\phi)$ depend on the magnetic symmetry
\cite{JETP89}. In this paper we consider stray-field free
configurations $\psi = \varphi + \pi/2$ arising in cubic
helimagnets (crystal classes 23 ($T$) and 432 ($O$))
and in uniaxial ferromagnets with $n$22 ($D_n$)
symmetry ($n$ = 3, 4, 6) \cite{JETP89,Nature06}.
For these handed ferromagnets 
the part of the Dzyaloshinskii-Moriya 
energy responsible for the stabilization of skyrmions
(i.e. that with in-plane gradients) can be written 
as $\tilde{w}_D = D (\Lambda_{xz}^{(y)}-\Lambda_{yz}^{(x)})$
where the $D$ is the Dzaloshinskii constant \cite{JETP89}.
The total energy of the axisymmetric
string in the layer of thickness $L$ 
can be written in the following 
reduced form $W = 2 \pi A L \widetilde{w}$
\begin{eqnarray}
&\widetilde{w}& = \int_0^{\infty} \bigg[\left(\theta_{\rho}^2+\frac{\sin^2 \theta}{\rho^2} \right)
+ \frac{4 \kappa}{\pi} \left( \theta_{\rho}+\frac{\sin  \theta \cos \theta}{\rho} \right)
\nonumber\\
&+& \sin^2 \theta +(H/H_a)(1-  \cos \theta)
 +\widetilde{w}_d (\rho, L)/Q \Big]\rho d\rho.
\label{energy}
\end{eqnarray}
Here, we use a new spatial variable $\rho = r /x_0$ 
and characteristic parameters 
\begin{eqnarray}
 x_0 = \sqrt{\frac{A}{K}}, \ H_a = \frac{K}{M},
\ Q = \frac{K}{2\pi M^2}, \ \kappa = \frac{\pi D}{4\sqrt{AK}},
\label{parameters}
\end{eqnarray}
where $x_0$ is the Bloch wall thickness, $H_a$ is the anisotropy field,
$Q$ is the quality factor \cite{Hubert98}, and the parameter 
$\kappa$ describes the relative contribution of the DM energy.
For $\kappa >1$ chiral modulations
in form of helices or skyrmion lattices 
become equilibrium states in bulk magnets \cite{JETP89}).
The stray-field energy of the axisymmetric string $ \tilde{w}_d$ is 
derived by solving the corresponding magnetostatic problem
\cite{Tu71}: $ \tilde{w}_d (\rho, L ) =(1- 2 \sin^2(\theta/2) \Omega (\rho, L)) $
\begin{eqnarray}
\Omega(\rho, L)=  \frac{x_0}{L}\int_0^{\infty} (1 - \cos \theta) \Xi (\rho, \xi, x_0/L) \xi d \xi ,
\label{strayfield}
\end{eqnarray}
where $\Xi (\rho, \xi, x) = \int_0^{\infty} J_0 (\xi\nu) J_0 (\rho\nu) \left[1 - \exp{(-\nu x)}\right] d \nu$,
$J_0 (x)$ are the zero-order Bessel functions.
The equations minimizing the functional $W$ (\ref{energy})
with boundary conditions $\theta (0) = \pi$,
$\theta(\infty) = 0$
yields the magnetization profiles for skyrmions 
$\theta (r)$ (Fig. \ref{f1}) in the phase space
of the four control parameters,
$\kappa$, $H/H_a$, $Q$, $L/x_0$.

\begin{figure}
\includegraphics[width=7.8cm]{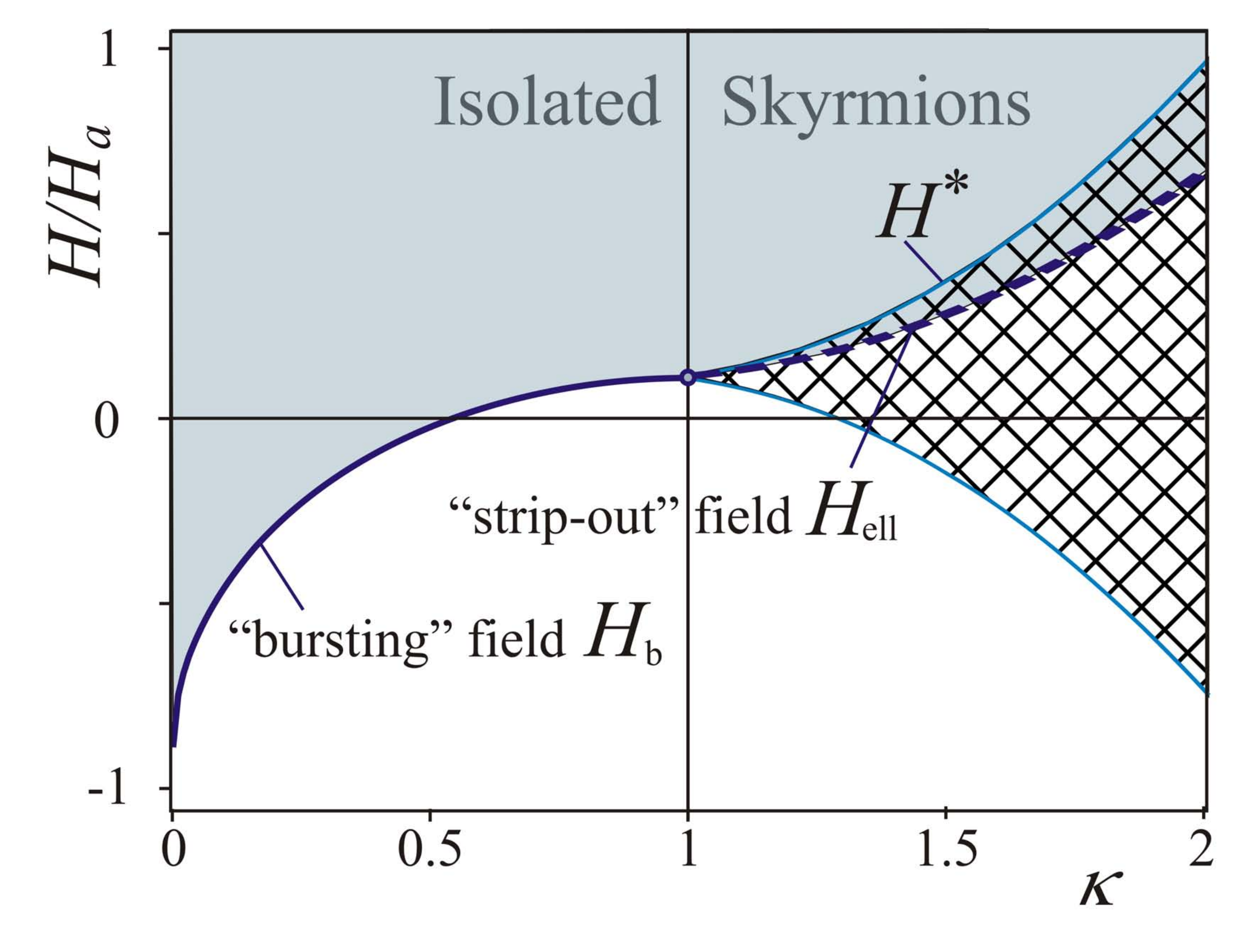}
\caption{ The magnetic phase diagram in
reduced variables $\kappa$ and 
applied magnetic field $H/H_a$ for fixed
values of $Q$ and the reduced layer thickness
$L/x_0$  indicate
the existence region of isolated skyrmions.
In the double-hatched area spatially modulated
phases (helicoids and skyrmion lattices)
correspond to the equilibrium state of
the film.
\label{diagram}
}
\end{figure}

Typical solutions $\theta(r)$
derived by numerical minimization 
of energy functional (\ref{energy})
are presented in Fig. \ref{profiles}.
In a broad range of the control parameters the skyrmion profiles
$\theta(r)$ consist of strongly localized arrow-like cores
with linear variation of the angle ($(\pi- \theta) \propto r$) 
and exponential "tails" with $ \theta \propto \kappa \exp{(-r/x_0)}$.
A skyrmion core radius can be defined as
\begin{eqnarray}
R_s = r_0 - \theta_0 (d \theta / d r)_0^{-1}.
\label{size}
\end{eqnarray}
where  $(r_0, \theta_0)$ are  the coordinates
of the inflection point $a$ and $(d \theta / d r)_0$
is the derivative in this point (Fig. \ref{f1}) \cite{JMMM99}.

%
%
%
The results in Fig. \ref{profiles}
demonstrate a variation of the skyrmion structure under 
the influence of magneto-dipole forces (parameters $Q$ and $L/x_0$).
This allows to adjust the skyrmion structure
and size by the variation of the
layer thickness $L$ or the value of the quality factor $Q$.

In the phase diagram in reduced parameters
$\kappa$, $H/H_a$ and with fixed values
of $Q$ and $L/x_0$ (Fig. \ref{diagram})
we indicate  the existence
area of isolated skyrmions.
Obviously skyrmions exist even in very high 
fields without collapse. 
At low fields the existence region of skyrmions 
is bound by several critical lines (Fig. \ref{diagram}).
They remain stable in zero and negative field
for $\kappa < 1$.
At a certain critical 
"bursting" field $H_{\textrm{b}}$
the skyrmion cores expand into
a homogeneous state with magnetization 
parallel to the applied field.
For $\kappa > 1$ skyrmions either condense
into lattices on the transition line $H^*$ or strip
out into a helical structure at 
a lower field $H_{\textrm{ell}}$.
All these exceptional 
features of chiral skyrmions 
rely on the {\em topological and energetic} stabilization 
of their core structure by chiral DM couplings.
Therefore, chiral skyrmions are fundamentally 
different from cylindrical {\em bubble} domains \cite{Hubert98}, 
which are intrinsically unstable and only arise
by the surface depolarization and the tension
of ordinary domain walls as an effect of the 
shape of a magnetized body.

\begin{figure}
\includegraphics[width=8cm]{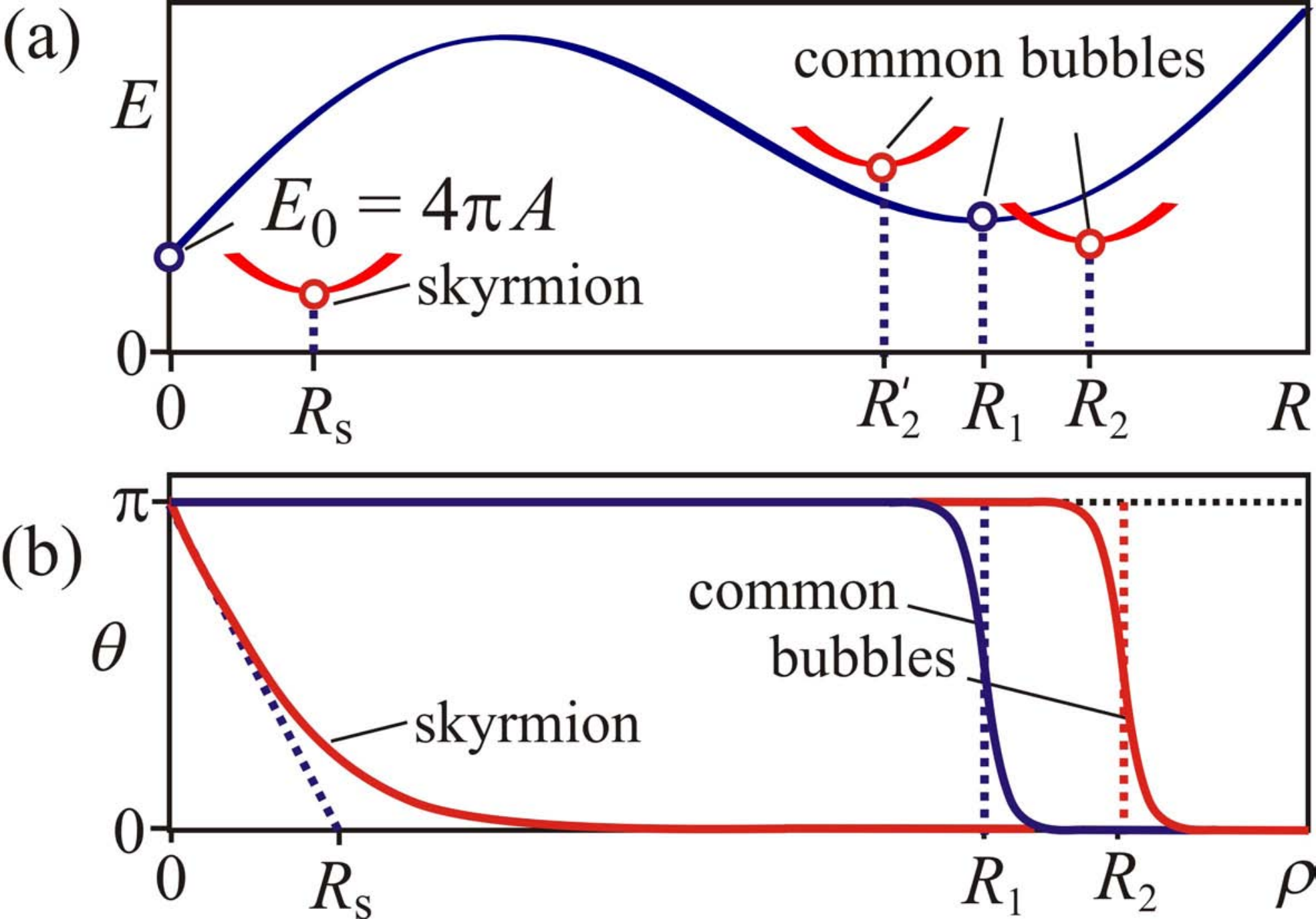}
\caption{ Energy $E$ of isolated bubbles and skyrmions
as a function of their sizes (a) and the corresponding
magnetization profiles $\theta (\rho)$) (b)
for a centrosymmetric magnet ($D = 0$) and that with 
finite Dzyaloshinskii-Moriya energy ($D \neq 0$).
\label{f2}}
\end{figure}

In Fig. \ref{f2} we demonstrate how axisymmetric solutions
for model (\ref{energy}) with $D = 0$ (thin (blue)
lines) transform into solutions with DM interactions
(thick (red) lines).
For $D = 0$  only solutions for cylindrical
domains (bubbles) exist as metastable
states in a certain range of parameters
$H/H_a$, $Q$, $L/x_0$ \cite{Tu71}.
Usually bubble profiles $\theta(r)$
consist of an extended core 
with $\theta = \pi$ separated by a thin domain
wall from the surrounding  homogeneously magnetized area 
with $\theta =0$ (Fig. \ref{f2} b) \cite{Hubert98,Thiele69,Tu71}.
In Fig. \ref{f2} $a$ energy (\ref{energy}) plotted
as a function of the bubble core size $E(r)$ 
(blue line) indicates a metastable solution for $r = R_1$.
Under the influence of DM interactions 
the profile of the energy density is modified 
and includes solutions for skyrmions with fixed rotation sense 
and finite radius $r = R_s$,
and solutions for isolated bubbles, 
which may have different rotation sense of the magnetization 
($r = R_2$ and $r = R'_2$) (red lines).
The coexistence of skyrmion and bubble solutions
occurs in a rather narrow range of the material
parameters. Outside this area bubbles are unstable
and only skyrmions  can arise in the film.

%
In order to to elucidate the physical mechanisms 
for the stabilization of chiral skyrmions 
in thin magnetic layers, a simplified 
semi-quantitative discussion is useful.
The profiles $\theta (r)$ in the skyrmion
center are linear. 
The strong localization 
of its core allows to use a linear ansatz 
$\theta = \pi (1 - r/R),  (0 < r < R)$,
$\theta = 0,  (r > R)$ as a suitable approximation
for skyrmion solutions.
With this trial function the equilibrium sizes of
the skyrmion are derived by minimization of 
the function 
\begin{eqnarray}
\Phi  = \widetilde{A} (H, Q) \nu^2 - 2 \kappa (x_0/L) \nu + \nu^3G(\nu)/Q , \ 
\nu = R/L,
\label{phi}
\end{eqnarray}
$G(\nu) = 2\int_0^{1} \cos (\pi \tau) \tau d \tau
\int_0^1 \cos^2 \left(\pi \xi/2\right)\Xi \left(\tau, \xi, \nu \right)  \xi d \xi$
with function $\Xi (\tau, \xi, x)$ introduced in Eq. (\ref{strayfield}),
$ \widetilde{A}(H,Q) = [1 + 8(1-4/\pi^2)(H/H_a)+(2/Q)(1-2/\pi^2)]$.
The linear term proportional to $\kappa$ is crucial 
for the existence of solutions with finite $R$.
The "stiffness" coefficient $\widetilde{A}$ includes uniaxial 
anisotropy, stray-field
and Zeeman energy contributions. 
Finally, the self dipole-dipole energy of the skyrmion
$G(\nu)$ introduces a dependence of its size 
on the layer thickness $L$ (Fig. \ref{profiles} Inset).

Recent experiments in nanolayers of magnetic materials 
with intrinsic (cubic helimagnets Fe$_{0.5}$Co$_{0.5}$Si 
and FeGe \cite{Yu10}) and with induced chirality 
(Fe/W bilayers) \cite{Heinze11}) report observations
of bound skyrmion states (lattices) (regions with $\kappa > 1$
in Fig. \ref{diagram}).
Isolated skyrmions with $R_s \approx $ 45 nm 
have been observed in a Fe$_{0.5}$Co$_{0.5}$Si  nanolayer ($L = $ 20 nm)
in the applied field larger than the critical field $H^* \approx$ 50 mT
(Fig. \ref{diagram}) \cite{Yu10}.
By using experimental data from Refs. \cite{Yu10,Beille81}
we found that for this sample $\kappa = 1.75$ and $x_0$ = 16 nm.
Our model results corroborate 
the identification of the observed 
magnetization patterns as chiral skyrmions.

In conclusion, in magnetic layers with intrinsic or
induced DM interactions isolated skyrmions 
with well-defined sized can exist 
as \textit{regular} solutions of micromagnetic
equations in a broad range of the material
parameters (Figs. \ref{profiles}, \ref{diagram}).
Chiral skyrmions as particle-like spots of reverse
magnetization can be considered as 
the smallest conceivable micromagnetic configuration.
Importantly and in contrast to alternative and traditional 
storage technologies based on highly coercive or patterned media,
chiral skyrmions can exist in {\em magnetically soft} 
nanolayers even at zero field, when sufficiently strong DM 
interactions can be induced them.
In such materials, chiral skyrmions can 
be easily induced, transported, and 
controlled, e.g., by electric currents 
and applied magnetic fields. 
Stable chiral skyrmions in extended layer systems
are promising objects for novel types of magnetic 
data storages, but also for logical bit-wise operations 
in extended layer systems.

\begin{acknowledgments}
The authors thank 
G. Bihlmayer,
S. Bl\"ugel, 
M. Bode, 
A.A. Leonov, and 
H. Wilhelm 
for helpful discussions.
\end{acknowledgments}



\begin{thebibliography}{99}

\bibitem{Dz64}  
I.\ E.\ Dzyaloshinskii, 
Sov.\ Phys.\ JETP {\textbf{19}}, 960 (1964).
%
\bibitem{JETP89}  
A.N.\ Bogdanov and D.A.\ Yablonsky,
Zh. Eksp. Teor. Fiz. {\textbf 95}, 178 (1989)
[Sov. Phys. JETP 68, 101 (1989)];
%
A. Bogdanov and A. Hubert,
J. Magn. Magn. Mater. {\textbf{138}}, 255 (1994).
%
\bibitem{Nature06} 
U.K. R\"o\ss ler et al.,
Nature \textbf{442}, 797 (2006),
%
U.K. R\"o\ss ler et al.,
J. Phys.: Conf. Ser. In press (2011)
(see also arXiv:1009.4849).

\bibitem{Kiselev10}
N.S. Kiselev et al.,
Phys. Rev. B {\textbf{81}}, 054409 (2010);
Appl. Phys. Lett., {\textbf{91}}, 132507 (2007);
{\textbf{93}}, 162502 (2008);
C. Bran et al., 
Phys. Rev. B {\textbf{79}}, 024430 (2009).

\bibitem{Schneider00}
M. Schneider et al., 
Appl. Phys. Lett., \textbf{77}, 2909 (2000);
A. Wachowiak et al., 
Science \textbf{298}, 577 (2002);
A. B. Butenko et al., 
Phys. Rev. B {\textbf{80}}, 134410 (2009).

\bibitem{PRL01}
A. N. Bogdanov, U. K. R\"o\ss ler,
Phys. Rev. Lett. {\textbf{87}}, 037203 (2001).

\bibitem{Yu10}
X. Z. Yu et al.,
Nature, {\textbf{465}}, 901 (2010);
Nature Mat. \textbf{10}, 106 (2011).


\bibitem{Heinze11}
 S. Heinze et al., 
 submitted to Nature Physics (2011)
 (see also APS March Meeting 2010, 
March 15-19,2010, abstract L34.014).

\bibitem{Hubert98}
A. Hubert, R. Sch{\"a}fer, {\it Magnetic Domains} 
(Springer, Berlin, 1998).

\bibitem{JMMM99}
A. Bogdanov and A. Hubert,
J. Magn. Magn. Mater. {\textbf{195}}, 182 (1999);
phys. stat. sol. (b) 
 {\textbf{186}}, 527 (1994).

\bibitem{Thiele69}
A. A. Thiele,
J. Appl. Phys. 
{\textbf{41}}, 1139 (1970);
Bell System Tech. Journal
{\textbf{48}},3287 (1969).

\bibitem{Tu71}
Y. Tu,
J. Appl. Phys. 
{\textbf{42}},5704 (1971);
W. J. DeBonte,
J. Appl. Phys. 
{\textbf{44}},1793 (1973).

\bibitem{Beille81}
J. Beille et al., 
J. Phys. F, {\textbf{11}}, 2153 (1981).

\end{thebibliography}
\end{document}